\documentclass[12pt]{article}

\advance\voffset by -2.0cm \advance\hoffset by -1.25cm
\usepackage{amsmath}
\usepackage{amssymb}
\usepackage{amsthm}
\usepackage{amsfonts}

\numberwithin{equation}{section}

\theoremstyle{definition}

\newcommand{\CC}{\mathbb{C}} 
\newcommand{\RR}{\mathbb{R}} 
\newcommand{\QQ}{\mathbb{Q}} 
\newcommand{\ZZ}{\mathbb{Z}} 
\newcommand{\PP}{\mathbb{P}} 

\DeclareMathOperator{\Sym}{Sym} 

\newcommand{\I}{\mathcal{I}}
\newcommand{\Z}{\mathcal{Z}}
\newcommand{\N}{\mathcal{N}} 
\newcommand{\J}{\mathcal{J}} 

\newcommand{\M}{{\mathcal M}} 
\newcommand{\tp}{\,{}^t\!} 
\newcommand{\be}{\begin{equation}}
\newcommand{\ee}{\end{equation}}

\def\perm{{\cal P}}  
\def\sprod#1#2{#1\cdot #2}    
\def\1{\frak 1}
\def\2{\frak 2}
\def\3{\frak 3}
\def\n{\frak n}
\def\m{\frak m}

\newlength{\oldcolsep}

\begin{document}

\title{Modular Invariant Regularization of String Determinants and the Serre GAGA Principle}
\author{Marco Matone}\date{}

\maketitle

\begin{center} Dipartimento di Fisica e Astronomia ``G. Galilei'' and Istituto
Nazionale di Fisica Nucleare \\
Universit\`a di Padova, Via Marzolo, 8-35131 Padova,
Italy\end{center}

\bigskip

\begin{abstract}
\noindent Since any string theory involves a path integration on the
world-sheet metric, their partition functions are volume forms on
the moduli space of genus $g$ Riemann surfaces  $\M_g$, or on its
super analog. It is well known that modular invariance fixes strong
constraints that in some cases appear only at higher genus. Here we
classify all the Weyl and modular invariant partition functions
given by the path integral on the world-sheet metric, together with
space-time coordinates, $b$-$c$ and/or $\beta$-$\gamma$ systems,
that correspond to volume forms on $\M_g$. This was a long standing
question, advocated by Belavin and Knizhnik, inspired by the Serre
GAGA principle and based on the properties of the Mumford forms.
The key observation is that the Bergman reproducing
kernel provides a Weyl and modular invariant way to remove the point
dependence that appears in the above string determinants, a property
that should have its superanalog based on the super Bergman
reproducing kernel. This is strictly related to the properties of
the propagator associated to the space-time coordinates. Such
partition functions $\Z[\J]$ have well-defined asymptotic behavior and can be considered
as a basis to represent a wide class of string theories. In particular, since
non-critical bosonic string partition functions $\Z_D$ are volume forms on $\M_g$, we suggest that
there is a mapping, based on bosonization and degeneration techniques, from the Liouville sector to first order systems that may
identify $\Z_D$ as a subclass of the $\Z[\J]$.
The appearance of $b$-$c$ and $\beta$-$\gamma$ systems of any conformal
weight shows that such theories are related to $W$ algebras. The
fact that in a large $N$ 't Hooft-like limit 2D $W_N$ minimal models
CFTs are related to higher spin gravitational theories on ${\rm
AdS}_3$, suggests that the string partition functions introduced
here may lead to a formulation of higher spin theories in a string
context.
\end{abstract}

\newpage

\section{Introduction}

Despite the great progress in string theory, finding a finite one in four dimensions is still an open question. Essentially, there is one candidate, superstring theory
but it must be compactified. Other possible string theories are those in non critical dimensions. However,
such theories are elusive; in essence, the proper way to treat the Liouville measure is still unknown.

Each string theory involves a path integration on the world-sheet
metric so that the corresponding partition functions are volume
forms on the moduli space of genus $g$ Riemann surfaces $\M_g$.
Other formulations involve super Riemann surfaces, so that the
partition functions correspond to volume forms on the supermoduli
space of super Riemann surfaces. However, at least for low genus, it
has been shown that there is a projection to a volume form on
$\M_g$. Other formulations of superstring theories suggest that it
may exist a mechanism, possibly involving a
rearrangement of the elementary fields, that may project the theory to
$\M_g$ even in higher genus.

One of the main results here is the classification of all the
possible Weyl and modular invariant string partition functions given
by the path integral on the world-sheet metric, together with space-time
coordinates, $b$-$c$ and/or $\beta$-$\gamma$ systems, that
correspond to volume forms on $\M_g$. This also provides the way to
test modular invariance that, as well known, fixes strong
constraints that may appear only at higher genus. It should be
stressed that the classification may also lead to uncover new
symmetries underlying string theories. The investigation is based on
two key properties of the Bergman reproducing kernel, namely its
Weyl and modular invariance. In this way, it is possible to remove
the point dependence that appears in the string determinants.

The role of modular invariance has been first noticed in \cite{Shapiro:1972ph}.
The Polyakov formulation of string theories \cite{Polyakov:1981rd} led to a considerable
progress in the covariant calculations of string partition functions and amplitudes
\cite{Alvarez:1982zi}-\cite{DiVecchia:1987uf} where modular invariance is a basic issue.

The need of classifying the string partition functions corresponding
to volume forms on $\M_g$ is also suggested by the strictly related
approaches to investigate superstring perturbation theory. The first
one is to consider the original theory trying to derive, step by
step from first principles, its measure on $\M_g$ or on the
supermoduli space. This is essentially due to D'Hoker and Phong in
the case of genus two \cite{D'Hoker:2001zp}. Very recently Witten
has proposed a systematic approach to the formulation on supermoduli
of superstring perturbation theory \cite{Witten:2012bg}.

Another approach is to change the elementary fields of the theory
and, again, derive the corresponding measure on $\M_g$. This is
mainly due to the Berkovits pure spinor formulation
\cite{BerkovitsFE}\cite{geometrica}.

Another way is to match the natural constraints of the theory with
the constraints imposed by the geometry of $\M_g$.  This has been
used to guess the form of the four-point function at arbitrary
genus,  leading to a result \cite{Matone:2005vm} which is also in
agreement with more recent investigations related to the $R^4$
non-renormalisation theorem in $N=4$ supergravity
\cite{Tourkine:2012ip}.  A similar ideology led to guess the
structure of the NSR partition function at any genus
\cite{D'Hoker:2001zp}\cite{Cacciatori:2007vk}, culminating with the
Grushevsky ansatz \cite{Grushevsky:2008zm}.

Of course the above  is a schematic view, as the three approaches
are strictly related, and each of them contributes to the other.

{}From the above lessons we learn that the basic physical motivation
to consider the possible volume forms on $\M_g$ is modular
invariance. It is just modular invariance that implies that the
string partition functions correspond to a volume form
$\M_g$. In this respect,
considering the case of higher genus Riemann surfaces is an
essential ingredient to understand the structure of a given theory.
There are other important issues, for example the problem of
treating the zero mode insertions in the path integral is quite
different in the case of the sphere and the torus with respect to
the case of negatively curved Riemann surfaces. As a consequence,
several questions, such as the one of modular invariant
regularization of the standard combination of string determinants,
may not appear in genus zero and one. This is essentially due to
Riemann-Roch theorem telling us that the space of zero modes of a
conformal field of a given weight may be zero dimensional on the
sphere and nontrivial for $g\geq2$. Similarly, since the torus is
flat, it is a special case as all the zero modes essentially
correspond to the constant.

Another reason to study volume forms on $\M_g$ for $g\geq 2$ goes
back to the Friedan-Shenker analytic approach to 2D CFT's
\cite{Friedan:1986ua}. Modular invariance is again the key issue.
Explicit examples are the ones by Gaberdiel and Volpato
\cite{Gaberdiel:2009rd}. They have shown that higher genus vacuum
amplitudes of a meromorphic conformal field theory uniquely
determine the affine symmetry of the theory. In particular, the
vacuum amplitudes of the $E_8\times E_8$ theory and the ${\rm
Spin}(32)/\ZZ_2$ theory differ at genus 5. The fact that the
discrepancy only arises at rather high genus is just a consequence
of the modular properties of higher genus amplitudes.

Another explicit realization of the Friedan-Shenker approach
concerns just the NSR superstring. In particular, it has been shown
in \cite{Matone:2010yv} that there exists a natural choice of the
local coordinate at the node on degenerate Riemann surfaces that
greatly simplifies the computations. This makes clear the power of
such an approach as now one
 may derive, at any genera, consistency relations involving the amplitudes and the measure. As a result
chiral superstring amplitudes can be obtained by factorizing the
higher genus chiral measure induced by considering suitable
degeneration limits of Riemann surfaces. Even in such investigations
modular invariance is the key symmetry.

Classifying string partition functions corresponding to volumes
forms on $\M_g$ may also lead to uncover new symmetries. String
theories essentially concern the bosonic and supersymmetric ones. The first is affected by the tachyon,
whereas the superstring, although free of such singularities, still
needs more than four dimensions and one has to compactify the extra
dimensions. In principle, it may happen that there are other string
theories with some underlying hidden symmetry. Investigating such a question requires the preliminary basic step
of classifying all forms on $\M_g$ satisfying the main properties that
a string theory should have. Let us summarize them.

\begin{enumerate}

\item Since each string theory involves the path integration over
the world-sheet metric, it should be a modular invariant $(3g-3,3g-3)$ form, i.e. a volume form on $\M_g$.

\item Such forms should correspond to determinants of laplacians
associated to the space-time coordinates to $b$-$c$ and/or $\beta$-$\gamma$ systems of any conformal weight.

\item The combination of such determinants should be Weyl invariant.

\end{enumerate}

\noindent Satisfying such conditions is essentially equivalent to require that the partition functions

\be
\int_{\M_g}{\cal Z}[\J]=\int DgDXD\Psi \exp(-S[X]-S[\Psi]) \ ,
\label{formidabile}\ee
correspond to volume forms on $\M_g$. Here $S[X]$ is the Polyakov action in
\be
D=26+2\sum_{k\in \I}n_kc_k \ ,
\label{puntiint}\ee
dimensions, where $c_k=6k^2-6k+1$ is (minus) 1/2 the central charge
of the non-chiral ($b$-$c$) $\beta$-$\gamma$ system of weight $k$.
$\I$ is the set of conformal weights $k\in\QQ$, $\J$ the set of
$n_k\in\ZZ/2$. $D\Psi$ denotes the product on $k\in \I$ of $|n_k|$
copies of the non-chiral measures, including the zero mode
insertions, of weight $k$ $b$-$c$ systems for $n_k>0$, or
$\beta$-$\gamma$ systems for $n_k<0$. $S[\Psi]$ is the sum of the
corresponding non-chiral $b$-$c$ and $\beta$-$\gamma$ actions. We will see that there exists a Weyl and modular invariant regularization of string determinant that eliminates the points dependence due to the insertion of zero modes. This will lead
to a consistent definition of ${\cal Z}[\J]$ as volume form on $\M_g$. One
of the main consequences of the present investigation is that the partition functions $\Z[\J]$
include a class of finite strings, even in 4D \cite{Matone:2012rw}.

The content of the paper is as follows. In Sec. 2 we review the partition function of $b$-$c$ and $\beta$-$\gamma$ systems. In particular, we
will consider the problem of treating the point dependence due to the zero mode insertions.
In Sec. 3 we will introduce the way to eliminate the point dependence of the zero mode insertions in the path integral, preserving
the modular and Weyl symmetries.
In Sec. 4 we will consider the bosonic string partition function $\Z_D$ in non-critical dimension $D$. In particular, since $\Z_D$
is a volume form on $\M_g$, its behavior at the boundary of $\M_g$ fixes some conditions that may be reproduced by $\Z[\J]$ for some $\J$. We suggest that
there is a map from the non gaussian measures on diffeomorphisms and the Liouville fields to the gaussian one that leads, via bosonization techniques, to represent the
Liouville sector by means of first-order systems. We will also show that $W$ algebras naturally arise in our construction. Interestingly, this may lead to
represent higher spin fields in a string context.

Although the prescription introduced in Sec. 3
is essentially the only well-defined recipe for any Riemann surface, there is a related approach which is defined on
canonical curves. These curves are the ones of genus two and the non-hyperelliptic compact Riemann surfaces with $g>2$.
In Sec. 5 we show that instead of integrating with $B^{1-n}(z_j,\bar z_j)$ each pair $b(z_j)\bar b(z_j)$ of the zero mode insertions, one may divide
them by the determinant of $B^{(n)}(z_j,\bar z_k)$, denoting the $n$-fold Hadamard product of
$B(z_j,\bar z_k)$. This
implies that the ratio of determinants of laplacians corresponding to the path integral on the
world-sheet metric, together and on space-time coordinates and to $b$-$c$
and/or $\beta$-$\gamma$ systems, become volume forms on the moduli space of canonical curves $\hat\M_g$. We
will show that  $\det B^{(n)}(z_j,\bar z_k)$ is expressed
in terms of the recently introduced vector-valued Teichm\"uller
modular forms \cite{Matone:2011ic}. We will also consider the Chern classes associated to our construction and see their relation with the tautological classes arising in 2D topological gravity.

Sec.  6 is devoted to further developments and to the conclusions. In the Appendix we introduce the mapping to the single indexing used in Section 5.

\section{Partition function of first order systems}

\subsection{Some notation}

Let $C$ be a Riemann surface of genus $g\geq2$ and denote by $\{\alpha_1,\ldots,\alpha_g,\beta_1,\ldots,\beta_g\}$ a symplectic basis of $H_1(C,\ZZ)$. Let
$\{\omega_i\}_{1\le i\le g}$ be the basis of $H^0(K_C)$ with the standard normalization $\oint_{\alpha_i}\omega_j=\delta_{ij}$ and
$\tau_{ij}=\oint_{\beta_i}\omega_j$ the Riemann period matrix. Set
$\tau_2\equiv{\rm Im}\,\tau$.
 The basis of $H_1(C,\ZZ)$ is determined up
to the
transformation
$$\begin{pmatrix}\alpha\\
\beta\end{pmatrix}\mapsto \begin{pmatrix}\tilde\alpha\\
\tilde\beta\end{pmatrix}=\begin{pmatrix}D & C \\ B& A\end{pmatrix}
\begin{pmatrix}\alpha\\ \beta\end{pmatrix}\ ,\qquad\qquad \gamma\equiv\begin{pmatrix}A
& B \\ C & D\end{pmatrix}\in \Gamma_g \ ,
$$
which induces the
following transformation on the period matrix
$$
\tau
\mapsto \gamma\cdot\tau=(A\tau+B)(C\tau+D)^{-1}\ .
$$
We will also consider the Deligne-Mumford
compactification of the moduli space of genus $g$ stable curves with $n$-punctures
$\overline{\M}_{g,n}$. It turns out that $\overline{\M}_{g}$
is a projective
variety with compactification
divisor
\be{D}=\overline{\cal M}_{g}\backslash {\cal M}_g={D}_0,\ldots,{D}_{[g/2]}\ .
\label{bobmarleys}\ee
A curve belongs to
${D}_{k>0}\cong \overline{\cal M}_{g-k,1}\times
\overline{\cal M}_{k,1}$ if it
has one node separating it into two components of genus $k$
and $g-k$. The locus in ${D}_0\cong \overline{\cal M}_{g-1,2}$
consists of surfaces that become, on removal of the node,
genus $g-1$ double punctured surfaces.
Surfaces with multiple nodes lie in the intersections
of the $D_k$.

The compactified moduli space $\overline{\cal M}_{g,n}$ of
stable curves with $n$-punctures
 is defined in an analogous way to $\overline{\cal M}_g$.
The important point now is
that the punctures never collide with the node.
In particular, the configurations of two colliding punctures are
stabilized by considering them as the limit in which the $n$-punctured
surface degenerates into a $(n-1)$-punctured curve and the thrice
punctured sphere.
Consider the Riemann theta function with characteristics
$$\theta \left[^a_b\right]\left(z,\tau\right)=
\sum_{k\in {\ZZ}^g}e^{\pi i \tp{(k+a)}\tau(k+a)+ 2\pi i \tp{(k+a)}
(z+b)} \ ,
$$ where $z\in \CC^g$ and $a,b\in{\RR}^g$.
If $\delta',\delta''\in\{0,1/2\}^g$, then $\theta
\left[\delta\right]\left(z,\tau\right):=\theta
\left[^{\delta'}_{\delta''}\right]\left(z,\tau\right)$ has definite
parity in $z$ $\theta \left[\delta\right]
\left(-z,\tau\right)=e(\delta) \theta \left[\delta\right]
\left(z,\tau\right)$, where $e(\delta):=e^{4\pi i\!\tp{\delta'}
\delta''}$. There are $2^{2g}$ different characteristics of definite
parity.
By Abel Theorem each one of such characteristics determines
the divisor class of a spin bundle $L_\delta\simeq K^{1\over2}_C$,
so that we may call them spin structures. There are $2^{g-1}(2^g+1)$
even and $2^{g-1}(2^g-1)$ odd spin structures.
Let $\nu$ be a non-singular odd characteristic. The holomorphic
1-differential
$$
h^2_\nu(p)=\sum_{1}^g\omega_i(p)
\partial_{z_i}\theta\left[\nu\right](z)_{|_{z=0}}\ ,$$ $p\in C$,
has $g-1$ double zeros. The prime form
$$
E(z,w)={\theta\left[\nu\right](w-z,\tau)\over
h_{\nu}(z)h_{\nu}(w)}\ ,
$$ is a holomorphic section of a line bundle
on $C\times C$, corresponding to a differential form of weight
$(-1/2,-1/2)$ on $\tilde C\times \tilde C$, where $\tilde C$ is the
universal cover of $C$. It has a first order zero along the diagonal
of $C\times C$. In particular, if $t$ is a local coordinate at $z\in
C$ such that $h_\nu=dt$, then
$$
E(z,w)={t(w)-t(z)\over\sqrt{dt(w)}\sqrt{dt(z)}}(1+{\cal O}((t(w)-t(z))^2)) \ .
$$
Note that $I(z+\tp\alpha n+\tp\beta m)=I(z)
+n+\tau m$, $m,n\in\ZZ^g$, and
$$
E(z+\tp\alpha n+\tp\beta m,w)=\chi e^{-\pi i \tp m \tau m -2\pi i \tp mI(z-w)}E(z,w) \ ,
$$
where $\chi=e^{2\pi i(\tp \nu' n- \tp \nu'' m)}\in\{-1,+1\}$, $m,n\in\ZZ^g$.
We will also consider the prime form $E(z,w)$ and the multivalued
$g/2$-differential $\sigma(z)$ on $C$ with empty divisor, satisfying
the property
$$
\sigma(z+\tp\alpha n+\tp\beta m)=\chi^{-g}e^{\pi i(g-1)\tp m \tau
m+2\pi i \tp m {\mathcal{K}}^z}\sigma(z)\ ,
$$
where $\chi$ is and ${\mathcal{K}^z}$ the vector of Riemann constants.
Such conditions fix $\sigma(z)$ only up to a factor independent of
$z$; the precise definition, to which we will refer, can be given,
following \cite{FayMAM}, on the universal covering of $C$  (see also
\cite{Matone:1900zz}).

\subsection{String determinants and Mumford forms}

Consider the covariant derivative $\nabla_z^{-n}$ acting on $-n$-differentials and
its adjoint $\nabla^z_{1-n}$. If  $\rho\equiv 2g_{z\bar z}$ is the metric tensor in local complex coordinates, that is $ds^2=2g_{z\bar z}dz d \bar z$, then
$$
\nabla_z^{-n}=\rho^{-n}\partial_z\rho^n \ , \qquad \nabla^z_{1-n}=\rho^{-1}\partial_{\bar z} .
$$
We will consider the determinants of such operators and  of the laplacian $\Delta_{1-n}=\nabla_z^{-n}\nabla_{1-n}^z$ acting on $1-n$ differentials.
Set
$$
c_n=6n^2-6n+1 \ .
$$
Let $\phi_1^n,\ldots,\phi_{N_n}^n$, $N_n=h^0(K_C^n)$, be a basis of $H^0(K_C^n)$. The partition function of the non-chiral $b$-$c$ system is \cite{Verlinde:1986kw}
$$
|\det \phi^n_j(z_k)|^2{\det'\Delta_{1-n}\over \det {\cal N}_n}=\int Db D\bar b Dc D\bar c\prod_{j}b(z_j)\bar b(z_j) e^{-{1\over2\pi}\int_C\sqrt g b\nabla^z_{1-n}c+c.c.}
$$
\be
=|Z_n(z_1,\ldots,z_{N_n})|^2e^{-c_n S_L(\rho)} \ ,
\label{nonchiraldetm}\ee
where $S_L(\rho)$ is the Liouville action and
$$
(\N_n)_{jk}=\int_C \bar\phi_j^n \rho^{1-n} \phi_k^n \ .
$$
Multiplying (\ref{nonchiraldetm}) by $\prod_1^{N_n}\rho^{1-n}(z_j,\bar z_j)$ and integrating over
$C^{N_n}$
by using the generalization of (3.3) in \cite{Matone:2005vm}, yields
\be
{\det}'\Delta_{1-n}=\int DbD\bar b Dc D\bar c \prod_1^{N_n}\int_{C^{N_n}} \rho^{1-n}(z_i,\bar z_i)b(z_i)\bar b(z_i)
e^{-{1\over2\pi}\int_C\sqrt g b\nabla^z_{1-n}c+c.c.} \ ,
\label{ecco}\ee
where we absorbed in the measure a numerical factor.
It turns out that for $n\neq1$ \cite{Verlinde:1986kw}
\be
Z_n(z_1,\ldots,z_{N_n})=Z_1[\omega]^{-1/2}\theta\Big(\sum_i z_i-(2n-1)\Delta\Big)\prod_{i<j}E(z_i,z_j)\prod_i\sigma^{2n-1}(z_i) \ ,
\label{zn}\ee
and
\be
Z_1(z_1,\ldots,z_g)=Z_1[\omega]^{-1/2}{\theta(\sum_i z_i-w-\Delta)\prod_{i<j}E(z_i,z_j)\prod_i\sigma(z_i)\over \sigma(z)\prod_iE(z_i,w)} \ .
\label{zuno}\ee
It turns out that
$$
Z_1[\omega]={Z_1(z_1,\ldots,z_g)\over\det\omega_j(z_k)} \ ,
$$
can be formally considered as the partition function of a chiral scalar. We have
$$
Z_1^{3/2}[\omega]={\theta(\sum_i z_i-w-\Delta)\prod_{i<j}E(z_i,z_j)\prod_i\sigma(z_i)\over \det\omega_j(z_k) \sigma(z)\prod_iE(z_i,w)} \ .
$$
Also note that
\be
Z_n[\phi^n]={Z_n(z_1,\ldots,z_{N_n})\over \det \phi^n_i(z_j)} \ ,
\label{zennennn}\ee
is independent of the points. One may easily check that the $g/2$-differential $\sigma$, the carrier of the gravitational anomaly, is missing in
\be
F_{g,n}[\phi^n]={Z_n[\phi^n]\over Z_1[\omega]^{c_n}} \ .
\label{star}\ee
This corresponds to the fact that such a ratio defines the Mumford form of degree $n$. More precisely, consider
the $(3g-2)$-dimensional complex space $\mathcal{C}_g$, called universal curve, built by placing over each point of $\M_g$ the corresponding curve $C$.
Consider the map $\pi$ projecting $\mathcal{C}_g$ to $\M_g$. Denote by  $L_n=R\pi_*(K^n_{\mathcal{C}_g/\mathcal{M}_g})$ the vector bundle on $\mathcal{M}_g$ of rank
$N_n$ with fiber $H^0(K_C^n)$ at the point of $\mathcal{M}_g$ representing $C$.
Let $\lambda_n=\det L_n$ be the determinant line bundle. The
Mumford isomorphism is \cite{Mumford}
$$
\lambda_n\cong\lambda_1^{\otimes c_n}\ .
$$
It turns out that, for each $n$, the Mumford form is \be
\mu_{g,n}=F_{g,n}[\phi^n]{\phi^n_1\wedge
\cdots\wedge\phi_{N_n}^n\over(\omega_1\wedge\cdots\wedge\omega_g)^{c_n}}
\ . \label{Mumfordformsss}\ee
Therefore, $\mu_{g,n}$ is the unique, up to a
constant, holomorphic section of $\lambda_n\otimes\lambda_1^{-c_n}$
nowhere vanishing on ${\cal M}_g$. The fact that the bosonic string
measure is essentially given by the Mumford form $\mu_{g,2}$ has
been first observed by Manin \cite{Manin:1986gx}. For $n=2$ its
expression in terms of theta functions has been given in
\cite{Manin:1986gx} whereas $\mu_{g,n}$ has been obtained in
\cite{Verlinde:1986kw}\cite{Belavin:1986cy}\cite{FayMAM}. See also
\cite{Catenacci:1986sy} for a related investigation.

\noindent By (\ref{nonchiraldetm}), (\ref{ecco}) and
(\ref{zennennn}) it follows that \be {\det}'\Delta_{1-n}=
|Z_n[\phi^n]|^2\det {\cal N}_ne^{-c_nS_L(\rho)} \ ,
\label{eccolodueee}\ee and the modulo square analog of
(\ref{Mumfordformsss}) is \be
{{\det}'\Delta_{1-n}\over({\det}'\Delta_0)^{c_n}}=|F_{g,n}[\phi^n]|^2{\det{\cal
N}_n\over({\cal N}_0\det{\cal N}_1)^{c_n}} \ . \label{analogodi}\ee

\subsection{$\beta$-$\gamma$ systems}

The above description extends to the case of $\beta$-$\gamma$ systems giving the inverse of the determinants with respect
to the case of the corresponding $b$-$c$ systems. Since
the treatment of zero modes is more subtle, let us shortly review the known results. Following \cite{Verlinde:1987sd}
and its notation, it turns out that the correlators of the chiral $\beta$-$\gamma$ system of weight $3/2$,
$$
G(y_i;w_j;x_k)=\int [D\beta D\gamma]_\delta e^{-S(\beta,\gamma)}\prod_{l=1}^m \gamma(y_i)\prod_{j=1}^{m-2g+2}\delta(\gamma(w_j))\prod_{k=1}^{m+1}\Theta(\beta(x_k)) \ ,
$$
can be expressed in terms of theta functions as
$$
G(y_i;w_j;x_k)={\prod_{l=1}^m Z_{{3\over2},\delta}(\sum x_k-\sum w_j-y_i-2\Delta)\over \prod_{l=1}^{m+1} Z_{{3\over2},\delta}(\sum_{k\neq l}x_k-\sum w_j -2\Delta)} \ ,
$$
where $Z_{{3\over2},\delta}$ denotes $Z_{3/2}[\phi^{3/2}]$ with now the theta function in (\ref{zn}) having characteristic $\delta$.
The chiral partition function corresponds to $m=2g-2$. In this case, taking $y_i=x_i$, $i=1,\ldots,2g-2$, such an expression reduces to
$$
G(x_i;0;x_i)={1\over
Z_{{3\over2},\delta}(\sum_1^{2g-2} x_i-2\Delta)} \ ,
$$
which is just the inverse of the corresponding
expression for the chiral $b$-$c$ system. Repeating the construction in the non-chiral case, and for any $n$, it can be seen that the
partition function for the non-chiral $\beta$-$\gamma$ system is just the inverse of (\ref{eccolodueee}).

\subsection{Modular invariance and zero modes}

Let us consider again Eq.(\ref{analogodi}). It shows that under a Weyl transformation
the ratio of laplacian determinants
$$
{\det'\Delta_{1-n}\over(\det'\Delta_0)^{c_n}} \ ,
$$
has the same transformation properties of
$${\det{\cal N}_n\over({\cal
N}_0\det{\cal N}_1)^{c_n}} \ .
$$
This means that the
anomalous transformation under Weyl rescaling of the two ratios is a kind of residual anomaly which follows from the
definition of the partition function. Actually, there is some degrees of freedom in treating
the zero modes, and one may also choose
$$ {{\det}'\Delta_{1-n}\over\det {\cal N}_n} \ ,
$$
rather than ${\det}'\Delta_{1-n}$.
However, this still gives a residual ambiguity, due to the
choice of the basis of the zero-modes
$\phi_1^n,\ldots,\phi_{N_n}^n$. To discuss such a question, it is
instructive to recall how the bosonic partition function in
the critical dimension is obtained.

\noindent First, the moduli part of the measure on the world-sheet metric in the path integral reduces to
$$
{{\det}'\Delta_{-1}\over\det {\cal N}_2}|\wedge^{max}\phi^2_j|^2 \ .
$$
Note that this is different from the partition function for a non-chiral $b$-$c$ system of weight 2, where $|\wedge^{max}\phi^2_j|^2$
is replaced by $|\det \phi_j^2(z_k)|^2$. By (\ref{eccolodueee})
$$
{{\det}'\Delta_{-1}\over\det {\cal N}_2}|\wedge^{max}\phi^2_j|^2= |Z_2[\phi^2]|^2e^{-13 S_L(\rho)}|\wedge^{max}\phi^2_j|^2 \ .
$$
The scalar integration gives
$({{\det}'\Delta_{0}/{\cal N}_0})^{-13}$
so that the critical bosonic string measure on $\M_g$ is
\be
\Z_{Pol}=\Bigg({{\det}'\Delta_{0}\over{\cal N}_0}\Bigg)^{-13}{{\det}'\Delta_{-1}\over\det {\cal N}_2}|\wedge^{max}\phi^2_j|^2=
\Bigg|{Z_2[\phi^2]\over Z_1[\omega]^{13}}\Bigg|^2
{|\wedge^{max}\phi^2_j|^2\over(\det{\cal N}_1)^{13}} \ .
\label{bosstrinmeas}\ee
Comparing such an expression with Eq.(\ref{star}) we get the precise relation with the Mumford form of degree 2
\be
\Z_{Pol}=|F_{g,2}[\phi^2]|^2{|\wedge^{max}\phi^2_j|^2\over(\det \tau_2)^{13}} \ ,
\label{precisina}\ee
where we used ${\cal N}_1=\tau_2$.

\section{String partition functions as volume forms on $\M_g$}

As discussed in the Introduction, since any possible string theory
would involve a path integration on the world-sheet metric, a
central question is to classify determinants of laplacians
associated to the space-time coordinates, $b$-$c$ and/or
$\beta$-$\gamma$ systems of any conformal weight, corresponding to
volume forms on $\M_g$. The combination of such determinants should
be Weyl and modular invariant.

Note that independence on the choice of the basis of $H^0(K^2_C)$
 in (\ref{bosstrinmeas}), and therefore the absence of a source
 of modular anomaly, is due to the fact that the metric measure
leads to a term $\det{\cal N}_2$ at the denominator, whose dependence
on the choice of the basis is balanced by
$|\wedge^{max}\phi^2_j|^2$. This means that apparently it is not
possible to define volume forms on $\M_g$ considering ratio of
laplacians of determinants, and therefore partition functions on the
world-sheet with $b$-$c$ systems, unless they come, as in the case
of the critical bosonic string, as an integration on metrics. In the
case of $n\neq2$, this would imply considering metrics on some
vector bundle. Let us explicitly illustrate the problem. According
to (\ref{nonchiraldetm}) and (\ref{analogodi}) the partition
function of $2c_n$ scalars and a $b$-$c$ system of weight $n$ would
give \be \Bigg({{\det}'\Delta_{0}\over{\cal
N}_0}\Bigg)^{-c_n}{{\det}'\Delta_{1-n}\over\det{\cal
N}_n}|\det\phi^n_j(z_k)|^2 =\Bigg|{Z_n[\phi^n]\over
Z_1[\omega]^{c_n}}\Bigg|^2{|\det\phi^n_j(z_k)|^2\over(\det{\cal
N}_1)^{c_n}} \ , \label{analogodibisse}\ee whose structure is
different from the one of the critical bosonic string
(\ref{bosstrinmeas}). For arbitrary $n$ the term
$|\wedge^{max}\phi^n_j|^2$ is replaced by $|\det\phi^n_j(z_k)|^2$.
They both guarantee independence from the choice of
the basis of $H^0(K_C)$, and therefore modular invariance.
On the other hand, it is commonly believed that removing the apparently harmless dependence on the points
in (\ref{analogodibisse}) may lead to a modular
or a Weyl anomaly. In this section we will show that this common belief is due to an undue identification between positive (1,1)-forms and path-integral metric.
In particular, depending on the context, the same positive definite (1,1)-form may correspond or not to the path-integral metric. Such an apparent ambiguity is quite evident
once one notes that the ratio of any two positive definite (1,1)-forms define a possible Weyl transformation, and this, of course, does not imply that all positive definite (1,1)-forms
should be Weyl transformed.

In the following we use a positive definite $(1,1)$-form to integrate on $C^{N_n}$ the zero mode insertions. It should
be observed that even if such a form has the same properties of a
metric, it is explicitly constructed in terms of Weyl and modular
invariant quantities. As a consequence, depending on the context, it
can be seen as a metric, so that getting the Weyl factor, or as a
Weyl invariant $(1,1)$-form. In this way the Weyl invariant ratios of regularized string determinants
correspond to $(0,0)$-forms on $\M_g$ that, multiplied by the Polyakov measure, define
volume forms on $\M_g$. Such a recipe is consistently defined on any Riemann surface.

\subsection{The fiber and the Weyl transformations}

Let $\I$ be the set of conformal weights $k\in\QQ$. Set $\J=\{n_k\in\ZZ/2|k\in \I\}$ and let
$D\Psi$ be the product on $k\in \I$ of $|n_k|$ copies of the
non-chiral measures, including the zero mode insertions, of weight
$k$ $b$-$c$ systems for $n_k>0$, or $\beta$-$\gamma$ systems for
$n_k<0$. We denote by $S[\Psi]$ the sum of the corresponding
non-chiral $b$-$c$ and $\beta$-$\gamma$ actions.

In (\ref{formidabile}) the Weyl invariant string partition functions corresponding to integrals of ${\cal Z}[\J]$ over $\M_g$ have been introduced.
The ${\cal Z}[\J]$
correspond to the Polyakov partition function $\Z_{Pol}$ times a
rational function of determinants of laplacians. Namely \be
\Z[\J]=\Z_{Pol}\prod_{k\in \I} \Z_k^{n_k} \ ,
\label{laprimaaaAAA}\ee
where, tentatively,
\be
\Z_n\sim\Bigg({{\det}'\Delta_{0}\over{\cal
N}_0}\Bigg)^{-c_n}{{\det}'\Delta_{1-n}\over\det{\cal
N}_n}|\det\phi^n_j(z_k)|^2 \ , \label{questionnn}\ee
with the right-hand side of (\ref{questionnn}) coinciding with the following
partition function \be \int DX Db D\bar b Dc D\bar
c{\prod_{j}b(z_j)\bar b(z_j)} \exp(-S[X] -{1\over2\pi}\int_C\sqrt g
b\nabla^z_{1-n}c+c.c.) \ , \label{otto}\ee
 where now $S[X]$ is the Polyakov action in
$2c_n$ dimensions.

If it were not for the dependence on the points, the right-hand side
of (\ref{questionnn}) would be the good definition for such a
combination of partition functions. The reason is that it satisfies
the condition $D=2c_n$ that guarantees the invariance of $\Z_n$ under Weyl
transformations
$$
g\longrightarrow e^\sigma g \ .
$$
As we will see, the precise definition of $\Z_n$ requires a modification of the
standard treatment of the zero modes that will lead to a point
independent, Weyl and modular invariant regularization of such ratio
of determinants.

In the previous section we saw that the string determinants are strictly related to the Mumford forms. In particular,
\be
|F_{g,n}[\phi^n]|^2=\Bigg({{\det}'\Delta_{0}\over{\cal N}_0\det{\cal
N}_1}\Bigg)^{-c_n}{{\det}'\Delta_{1-n}\over\det{\cal N}_n} \ ,
\label{primochechiamo}\ee Until now, the only Mumford form which
appeared in string theory is $\mu_{g,2}$, that is the one defining
the Polyakov measure. On the other hand,
to express
(\ref{laprimaaaAAA}) as well-defined quantities on $\M_g$, requires one
to find what is the precise correspondence between the modulo square
of Mumford forms $\mu_{g,n}$ and $\Z_n$. This has
been an open question since the times of the covariant formulation
of string theories. In particular, Belavin and Knizhnik stressed
that the Mumford forms are Weyl anomaly free \cite{Belavin:1986cy}.
They also observed that since the holomorphic structure of $\M_g$ is
an algebraic structure, it follows that any holomorphic quantity on
$\M_g$, such as $\mu_{g,2}$, is an algebraic object. This is
essentially a consequence of the Serre GAGA principle \cite{Serre}
that led to the following conjecture \cite{Belavin:1986cy}

\vspace{.6cm}

\noindent {\it Multiloop amplitudes (and not only vacuum amplitudes)
in any conformally invariant string theory (such as the bosonic in
$D=26$ or the superstring in $D=10$) can be expressed in terms of
algebraic objects (functions or sections of holomorphic bundles) on
the moduli space of Riemann surfaces. Quantum geometry is therefore
the complex geometry of the space $\overline\M_g$.}

\vspace{.6cm}

\noindent
In this context, it was suggested in \cite{Matone:1993nf} that
non-critical strings may be formulated in terms of Mumford forms. In spite of its geometrical and physical elegance, the Belavin-Knizhnik
conjecture has not been developed so far. There are several reasons for that.  We will see
that such reasons are strictly related and admit a natural physical solution leading to a modular invariant
regularization of the string determinants.

An obvious reason why apparently the Mumford forms of degree $n\neq2$ should not play a role
in string theory is that only $|\mu_{g,2}|^2$ defines a volume form on $\M_g$. This question was in debate during the eighties.
To map $|\mu_{g,n}|^2$ to volume forms on $\M_g$ requires solving the problem of the fiber, that is, loosely speaking
replacing the wedge products of $n$-differentials by scalar quantities. Let us recall where the point is.

First,
note that $|\wedge^{max}\phi_j^1|^2$ and  $\det \int_C\bar\phi_j^1\wedge\phi_k^1$ have the same modular transformations.
In particular,
the Hodge fiber $|\wedge^{max}\omega_i|^2$ maps to
$$
\det {1\over2i}\int_C\bar\omega_j\wedge\omega_k=\det \tau_2 \ ,
$$
where $\tau_2={\rm Im} \tau$, with
$\tau_{jk}=\oint_{\beta_j}\omega_k$ the Riemann period matrix.

Even the wedge products
$|\wedge^{max}\phi_i^2|^2$ appearing in the Polyakov string are not a problem; since they represent the
infinitesimal volume elements on $\M_g$, it is just the term coming
from the path integration on the metric. However, it is commonly believed that integrating on $C^{N_n}$ in the case of
$|\wedge^{max}\phi_i^n|^2$, $n\neq2$ (or even $n=2$, if one
wants to reduce $|\mu_{g,2}|^2$ to a scalar quantity) leads to a Weyl anomaly.
The reason is that  $|\det\phi_{j}^n(z_k)|^2$ requires a metric to be integrated. However, the metric with respect to which one has to consider the Weyl transformations is the one
on which one integrates in the path integral and these concern only the metric defining the laplacians and the associated zero mode matrices
$({\N_n})_{ij}$. This means that, in principle, one can multiply $|\det\phi_{j}^n(z_k)|^2$  by the product of any $(1-n,1-n)$ form in $z_k$ and then integrating over $C^{N_n}$ without
worrying about any Weyl anomaly. Nevertheless, it is clear that this would lead to a considerable ambiguity.
In this respect note that making a Weyl transformation requires one to identify
which ones of the $(1,1)$-forms in a given expression correspond to the metric or are defined in a metric dependent way.
The question then is to find a positive definite $(1,1)$-form which is defined in a Weyl and modular invariant way.

\subsection{Weyl and modular invariant integration}

A key observation is that since the kernel of
$\partial_{\bar z}$ is metric independent, it follows that the space of zero modes $H^0(K_C^n)$
is Weyl invariant.
We then introduce the $(1,1)$-forms
$$
B_Y(z,\bar z)= \sum_{1}^g \phi_j^1(z) Y_{jk} \bar\phi_k^1(z) \ ,
$$
and will integrate $\prod_{k=1}^{N_n}B_Y^{1-n}(z_k,\bar z_k)|\det\phi_{j}^n(z_k)|^2$ on $C^{N_n}$. To fix $Y$ we use modular invariance. To this end, we use the fact that the dependence on the modular transformations
of the integration on $C$ is entirely given by the transformation properties of the integrand. This means that $B_Y(z,\bar z)$ is modular invariant and positive definite if $Y^{-1}_{jk}={1\over2i}\int_C \phi_j^1\wedge\bar\phi_k^1$. With this choice, $B_Y(z,\bar w)$ coincides with
\be
B(z,\bar w)=\sum_{1}^g\omega_j(z)(\tau_2^{-1})_{jk}\bar\omega_k(w)  \ ,
\label{brk}\ee
which is the Bergman reproducing kernel \cite{FayMAM}.
It should be stressed that the above investigation does not imply that $B(z,\bar z)$ cannot be considered as a metric on $C$. This is a general fact that holds for any positive definite $(1,1)$-form. In particular, any ratio between two positive definite $(1,1)$-forms
defines a possible Weyl transformation. In other words, considering a reference metric tensor $\rho$, there exists a Weyl transformation such that $e^\sigma \rho=B$. On the other hand, this does not mean that under a Weyl transformation one should transform all possible $(1,1)$-forms. For example, if an expression contains the term $\rho B$, under the Weyl transformation $e^\sigma \rho=B$ would transform to $B^2$, whose inverse is $\rho B$.
We will perform a similar transformation that will simplify the expressions of ${\cal Z}_n$ (see Eq.(\ref{questionnnsolveddue})).

It is instructive to recall that the Bergman reproducing kernel also appears in the two point function of a scalar field
$$
f(z,w)=\langle X(z)X(w)\rangle \ .
$$
This is the Green function for the scalar laplacian, so that it satisfies the following equations
$$
\int_C \sqrt g f(z,w)=0 \ ,
$$
$$
\partial_z\partial_{\bar z} f(z,w)=-\pi\delta(z-w)+{\pi\over A}\sqrt{g(z)} \ ,
$$
$$
\partial_z\partial_{\bar w} f(z,w)=\pi\delta(z-w)-\pi B(z,\bar w)\ ,
$$
where $A=\int_C\sqrt g$.
This shows that the building block of the string correlation functions naturally selects a Weyl and modular invariant kernel,
providing another way to show that the Bergman reproducing kernel (\ref{brk})
is intrinsically defined, it depends only on the complex structure of $C$ and on the points $z$ and $w$.
One may say that it is essentially the only way to select two points on $C$ in a way which is anomaly free. In particular,
since $B(z,\bar z)$ is a positive definite $(1,1)$-form, it can be used to integrate the zero modes in a Weyl and modular invariant
way. Multiplying (\ref{nonchiraldetm}) by $\prod_1^{N_n}B^{1-n}(z_j,\bar z_j)$ and integrating over
$C^{N_n}$ leads to
\be
X_n:=\det
\M_n{\det '\Delta_{1-n}\over\det{\cal
N}_n} =\int Db D\bar b Dc D\bar
c(b\bar b)_n\exp(-{1\over2\pi}\int_C\sqrt g b\nabla^z_{1-n}c+c.c.) \ ,
\label{nneerr}\ee
where
$$
(b\bar b)_n= \int_{C^{N_n}} \prod_j B^{1-n}(z_j,\bar z_j)b(z_j)\bar b(z_j) \ .
$$
and
$$
(\M_n)_{jk}=\int_C \bar \phi_j^n(z) B^{1-n}(z,\bar z) \phi_k^n(z) \ .
$$
The string determinants are then
\be
\Z_n=X_n\int DX  e^{-S[X]}\ .
\label{Pnove}\ee
Furthermore, we now have the precise identification of the string determinants in (\ref{questionnn})
\be \Z_n=\Bigg({{\det}'\Delta_{0}\over{\cal
N}_0}\Bigg)^{-c_n} {\det
\M_n\det '\Delta_{1-n}\over\det{\cal
N}_n} \ .
\label{questionnnsolved}\ee
Note that
$\det\M_1=\det\N_1=\det\tau_2$.
It should be stressed that the above prescription is equivalent to map $|\mu_{g,n}|^2$
to the $(0,0)$-forms
\be
\Z_n=|F_{g,n}|^2{\det\M_n\over(\det \tau_2)^{c_n}} \ ,
\label{fgenne}\ee
which is equivalent to map the modulo square of the wedge products in the Mumford to $(0,0)$-forms, that is
$$
{|\wedge^{max}\phi_j^n|^2\over|\wedge^g\omega_j|^{2c_n}} \longrightarrow {\det\M_n\over(\det \tau_2)^{c_n}} \ .
$$
The above results show that if $D=2c_n$, then the corresponding partition function admits a natural Weyl and modular invariant regularization which is point independent.

There is a remarkable mechanism that simplifies the expression of $\Z_n$ in (\ref{questionnnsolved}). Namely, since $|F_{g,n}|^2$ in (\ref{primochechiamo}) is Weyl invariant, we can just choose
as metric the Bergman metric, that is
$$
\rho(z,\bar z)=B(z,\bar z) \ .
$$
In this way
$$
\det \N_n=\det \M_n \ , \qquad  \N_0=\int_CB=g \ .
$$
The result is that the basis $\phi_j^n$ does not appear at all and
$\Z_n$ assumes the simplified form \be \Z_n=({\det}'\Delta_{B,
0})^{-c_n} {\det}'\Delta_{B, 1-n} \ , \label{questionnnsolveddue}\ee
where the laplacians are defined with respect to the Bergman metric.
In particular, with this choice \be \Z[\J]=({\det}'\Delta_{B,
0})^{-13-\sum_{k\in \I}n_kc_k}{\det '\Delta_{B,-1}\over\det{\M}_2}
\prod_{k\in\I}({\det}'\Delta_{B,
1-n})^{n_k}|\wedge^{3g-3}\phi_j^2|^2\ , \label{PPformidabile}\ee
that, by (\ref{fgenne}), is \be \Z[\J]={|F_{g,2}|^2\over
(\det\tau_2)^{13}}\prod_{k\in\I} \Bigg({|F_{g,k}|^2\det\M_k\over
(\det\tau_2)^{c_k}}\Bigg)^{n_k}|\wedge^{3g-3}\phi_j^2|^2 \ ,
\label{dfyhg}\ee which, by (\ref{zn}), (\ref{zuno}) and (\ref{star})
provides the expression of $\Z[\J]$ in terms of theta functions.

\section{$\Z[\J]$ non-critical strings and $W$-algebras}

\subsection{Non critical strings}

The typical singularities of string theories arise when some handle
of the Riemann surface is pinched. Such a degenerate surface belongs
to the Deligne-Mumford boundary $\partial\overline\M_g$. The
standard example is the tachyon singularity. Let us consider the singularity structure
at $\partial\overline\M_g$ associated to the Mumford forms for any $n$. The tachyon singularity of the critical
bosonic string corresponds to $n=2$.

In \cite{FayMAM} Fay derived the singular behavior of the Mumford
forms at the Deligne-Mumford boundary. He used Bers-like basis
$\phi^n_{t}=\{\phi^n_{i,t}\}_{i\in I_{N_n}}$ for $H^0(K_C^n)$. It
turns out that in the case of separating degeneration \be
F_{g,n}[\phi^n_{t}]\sim t^{-n(n-1)/2}{E(a,b)^{n-n^2}\over
(2\pi i)^{(2n-1)^2}}F_{g-1,n}[\phi^n] \ , \label{everouno}\ee
where $a,b$ are two points identified on the smooth genus $g-1$
curve. In the case of degeneration corresponding to a reducible
singular curve obtained by identifying points on two smooth curves
of genus $g_1$ and $g-g_1$, we have \be
F_{g,n}[\phi^n_{t}]\sim \epsilon t^{-n(n-1)/2}
F_{g-g_1,n}[\phi^n]F_{g_1,n}[\phi^n,] \ ,
\label{everodue}\ee where $\epsilon$ is a fixed $(2g-2)$th root of
unity. The tachyon singularity of the critical
bosonic string corresponds to $n=2$.
The above asymptotic analysis is just a consequence of the Grothendieck-Riemann-Roch theorem and of the Mumford formula.

Let us consider the measure on the world-sheet metric. This includes the integration on the diffeomorphisms and on the Liouville field
$D_gv^z D_gv^{\bar z}D_g\sigma$. It is well known that such measures are not gaussian. This is the problem of quantizing
Liouville theory. Let us consider the measure on the diffeomorphisms.
Since
$$\langle v,v\rangle_{g=e^\sigma\hat g}=
\int_C \sqrt{\hat g}{\hat g_{ab}}e^{2\sigma}v^a v^b \ ,
$$
it follows that ${\rm Vol}_g(Diff(\Sigma))$
depends on $\sigma$.
In critical string theory it is assumed
that such a dependence can be absorbed into $D_g\sigma$ and then one
drops the $D_gv^z D_gv^{\bar z}$ term.
However for $D\ne 26$ such a procedure still needs to be fully understood.
To overcome such a question
we consider the bosonic partition function
in non critical dimensions
\be
\int_{\M_g}\Z_D=\int DgDX \exp(-S[X]) \ ,
\label{noncrit}\ee
where $S(X)$ is the Polyakov action in $D$ dimensions, so that $\Z_{Pol}=\Z_{26}$.
Of course, like $\Z[\J]$, even $\Z_D$ must be a well-defined volume form on ${\cal M}_g$.
Since $\Z_D$ should be a volume form on $\M_g$, the central charge of the Liouville sector is
\be
c_L=26-D \ .
\label{centralLiouville}\ee
This is the reason why (\ref{centralLiouville}) has the same structure of (\ref{puntiint}).
This suggests considering
\be
c_L=-\sum_{n_k\in \I}2n_k c_k \ .
\label{cunoo}\ee
By means of a semiclassical analysis
it should be possible to check the behavior of $\Z_D$ when the Riemann surface degenerates, that is at the Deligne-Mumford boundary $\partial \overline\M_g$. This fixes some condition
on $\J$ in such a way that $\Z[\J]$ has the same behavior of $\Z_D$. This would select the first order systems as possible candidates to represent the Liouville partition functions. This means that there is a mechanism,
related to the bosonization of first order systems,
mapping the non-gaussian measures to the gaussian ones of the $b$-$c$ and $\beta$-$\gamma$ systems, as suggested in \cite{Matone:1993nf}.

It is instructive to understand what happens in the case in which the Liouville sector can be represented by a single first order system. Let us first consider the
case of the $\beta$-$\gamma$ system. This means
\be
c_L=12 k^2-12 k+2 \ ,
\label{cielle}\ee
that is the weight of the corresponding $\beta$-$\gamma$ system is
\be
k={3\pm\sqrt{81-3D}\over6} \ .
\label{kbg}\ee
Rational values of $k$ with integer $D$ are obtained for
$$
D=0,15, 24, 27 \ ,
$$
corresponding to
$$
k=2, 3/2, 1, 1/2 \ .
$$
Similarly, in the case of $b$-$c$ systems one gets
\be
c_L=-12 k^2+12 k-2 \ ,
\label{cielledue}\ee
\be
k={3\pm\sqrt{3D-75}\over6} \ ,
\label{kbgb}\ee
and rational values of $k$ with integer $D$ correspond to
$$
D=25+3n^2 \ ,
$$
with
$$
k={1\over2}(n+1) \ ,
$$
$n=0,1,2,\ldots$.
The above can be generalized to real and even complex values of $n$.
From
the point of view of the $b$-$c$ and $\beta$-$\gamma$ systems this
is always possible just because the action contains terms such as
$b\bar\partial c$ and $\beta\bar\partial \gamma$, which are well-defined $(1,1)$ forms even for $n\in\CC$.
A related aspect has been considered in
\cite{Bonora:1989wk} where a
general method to absorb the spin fields in $b$-$c$ systems of real
weight was introduced. Subsequently, complex powers of line bundles in connection with
string scattering amplitudes have been considered by Voronov \cite{Voronov:1990pp}.
Such an extension of first order systems to real and, more generally, complex weights is of considerable interest and should be further investigated.

\subsection{$W$-algebras and volume forms on $\M_g$}

There is a nice interpretation of the Mumford forms that should be
further investigated. A reason why the Polyakov partition function
leads to $|\mu_{g,2}|^2$ is that the world-sheet metric is deformed
by the Beltrami differentials; these are the dual of $H^0(K_C^2)$. It follows that
$|\mu_{g,n}|^2$ should be associated to a theory containing the
path integration on a metric deformed by the generalized Beltrami
differentials introduced \cite{Matone:1993tj}. These are the dual of
$H^0(K^n_C)$. In particular, using the single indexing introduced in the Appendix, one may consider the map
$$\omega_i^{(n)}\mapsto{1\over2\pi i}d\tau_i^{(n)} \ ,
$$
that defines the tangent space to the moduli space associated to the
holomorphic $n$-differentials, that is the moduli space of vector
bundles on Riemann surfaces. This is like the Kodaira-Spencer map
sending $\omega_i^{(2)}$ to ${1\over2\pi i}d\tau_i^{(2)}$. It follows that $|\mu_{g,n}|^2$ should correspond
to
$$
\int Dg^{(n)}D\Phi e^{-S[\phi]} \ .
$$
More generally, one should consider partition functions such as \be
\int \prod_{k\in\I}Dg^{(k)}D\Phi e^{-S[\Phi]} \ ,
\label{suvectorbundles}\ee where $g^{(k)}$ are the metrics
associated to $H^0(K_C^k)$, whose dual spaces are the generalized
Beltrami differentials, and $S[\Phi]$ some conformal action leading
to an anomaly $2\sum_{k\in\I}c_k$. Understanding the field content
of (\ref{suvectorbundles}) should lead to formulate a class of
Conformal Field Theories. In this respect note that
the generalized Beltrami differentials are related to the chiral
split for the higher order diffeomorphism anomalies. The Wess-Zumino
conditions correspond to the cocycle identities (see Sec. 3.4 of
\cite{Matone:1993tj}).

Since such theories are associated to
$W$ algebras, it would be interesting to investigate a possible
relation with higher spin theories. In particular, note that in a
large $N$ 't Hooft-like limit 2D $W_N$ minimal models CFTs are
related to higher spin gravitational theories on ${\rm AdS}_3$
\cite{Gaberdiel:2012uj}. This may suggest the existence of a
formulation of higher spin theories in a string context.

\section{Volume forms on $\M_g$ and Hadamard product}

Another interesting possibility to use the Bergman
reproducing kernel to remove the point dependence due to the zero
mode insertions is to consider the determinant of the Hadamard
$n$-fold product of $B(z_i,\bar z_j)$. Although such a recipe may
lead, depending on the set $\J$, to zeroes or singularities on the
hyperelliptic Riemann surfaces, it defines volume forms on $\M_g$
with interesting properties.

\subsection{Hadamard product of the Bergman kernel}

Let us first shortly review a result in \cite{Matone:1900zz}. Here we use the single index notation
defined in the Appendix.
In the following we consider the matrix $B^{\circ n}(z_j,\bar z_k)$ whose $j,k$th entry is
$(B(z_j,\bar z_k))^n$. This is the $n$-fold Hadamard product of $B(z_j,\bar z_k)$.
Furthermore, we will consider the determinant of $B^{\circ n}(z_j,\bar z_k)$, with the indices $j,k$ ranging from
$1$ to $N_n$. For each positive integer $n$ defines $I_n=\{1,\ldots,n\}$.
For all $z_i,w_i\in C$, $i\in I_{N_n}$,
we define
\be
{K}_n={\det B^{\circ n}(z_i,\bar z_j)\over |\det\phi_j^n(z_k)|^2}\bigl|\kappa[\phi^n]\bigr|^2 \ .
\label{kenneaaa}\ee

\noindent
It can be proved that  \cite{Matone:1900zz}
\be
{K}_n=\sum_{{i_{N_n}>\ldots>
i_1=1\atop j_{N_n}>\ldots
> j_{1}=1}}^{M_n}\kappa[\omega^{(n)}_{i_1},\ldots,\omega^{(n)}_{i_{N_n}}]
{|\tau_2^{-1}\cdots \tau_2^{-1}|^{i_1\ldots i_{N_n}}_{j_1\ldots
j_{N_n}}\over
\prod_{k=1}^{{N_n}}\chi_{i_k}\chi_{j_k}}\,\bar\kappa[\omega^{(n)}_{j_1},\ldots,
\omega^{(n)}_{j_{N_n}}]\ ,
\label{effennea}\ee
where  $|A\cdots A|^{i_1\ldots i_m}_{j_1\ldots j_m}$
denotes the minors of $(A\cdots A)$
$$|A\cdots A|^{i_1\ldots i_m}_{j_1\ldots j_m}=\det_{{i\in {i_1,\ldots,i_m}\atop j\in
{j_1,\ldots,j_m}}}(A\cdots A)_{ij}\ ,$$ $i_1,\ldots,i_m,j_1,\ldots,j_m\in I_{M_n}$,
with $m\in I_{M_n}$.

\subsection{Zero modes, $\det B^{\circ n}(z_j,\bar z_k)$ and volume forms on $\M_g$}

In the following we investigate a way to absorb the point dependence due to the insertion of the zero modes which is related to the one introduced in Sec. 3.
This may lead to zeros or singularities on the hyperelliptic Riemann surfaces of genus greater than two. The moduli space of compact Riemann surfaces
that does not contain
such Riemann surfaces is called the moduli space of canonical curves $\hat\M_g$.

\noindent Let us set
$$
\kappa[\phi^1]={1\over Z_1[\phi^1]^{3\over2}}\ ,
$$
and, for $n>1$
$$
\kappa[\phi^n]={1\over Z_1[\omega]^{3\over2}Z_n[\phi^n]} \ .
$$
Note that $\det B(z_j,z_k)=|\det\omega_j(z_k)|^2/\det \tau_2$, so that
$$
K_1=
{|\kappa[\omega]|^2\over\det \tau_2} \ .
$$
Also note that
replacing $|\omega_1\wedge\cdots\wedge\omega_g|^2$ in $|\mu_{g,n}|^2$ by $\det \tau_2$ does not break modular invariance.
Together with the Kodaira-Spencer map $\omega_i\omega_j\to
d\tau_{ij}/(2\pi i)$, this is what one does in passing from the
Mumford form $|\mu_{g,2}|^2$ to the Polyakov measure. What is less obvious
is the analog of $\det \tau_2$ when one considers the wedge
products $\phi^n_1\wedge\cdots\wedge\phi^n_{N_n}$. First notice that
due to the term $\det\phi^n_i(z_j)$ in $\kappa[\phi^n]$, the
Mumford forms are independent of the choice of the bases
$\phi^n_1,\ldots,\phi^n_{N_n}$. On the other hand, using $\tau_{ij}$
as moduli parameters naturally requires one to use $N_n$ elements
$\omega_i^{(n)}$ as basis of $H^0(K_C^n)$, a fact that led to the
concept of vector-valued Teichm\"uller modular forms
\cite{Matone:2011ic}\cite{Matone:2012wy}. Therefore, one has to
consider
$\omega_{i_1}^{(n)}\wedge\cdots\wedge\omega_{i_{N_n}}^{(n)}$. In
order to have a volume form for $n=2$, one has to consider the
Kodaira-Spencer map\footnote{Note that such two terms have the same
transformation properties under ${\rm Sp}(2g,\ZZ)$.}
$$|\omega_{i_1}^{(2)}\wedge\cdots\wedge\omega_{i_{3g-3}}^{(2)}|^2 \longrightarrow |d\tau_{i_1}\wedge\cdots\wedge d\tau_{i_{3g-3}}|^2 \ .
$$
However, as it will be clear below, we can also consider a map involving $K_2$. Let us stress that,
thanks to the term $\det \omega^{(2)}_i(z_j)$ in the denominator of $\mu_{g,2}$, when $d\tau_{i_1}\wedge\cdots\wedge d\tau_{i_{3g-3}}$ vanishes in some subspace
of $\M_g$, e.g. in the hyperelliptic loci of genus $g\geq3$, this is balanced by the vanishing of $\det \omega^{(2)}_i(z_j)$.

\noindent
The situation is different when looking for the analog of the map
$|\omega_1\wedge\cdots\wedge\omega_g|^2\longrightarrow\det \tau_2$ in the case
of $\omega_{i_1}^{(n)}\wedge\cdots\wedge\omega_{i_{N_n}}^{(n)}$,
even in the case $n=2$. The answer is to replace the
modulo square of Mumford forms building blocks by
$K_n$. In particular,
$$ \Big|{\kappa[\omega]\over
\omega_1\wedge\cdots\wedge\omega_g}\Big|^2 \longrightarrow
K_1 \ ,
$$ and, for $n\neq1$
$$
\Big|{\kappa[\phi^n]\over
\phi^n_1\wedge\cdots\wedge\phi^n_{N_n}}\Big|^2 \longrightarrow
K_n \ ,
$$
so that
\be
|\mu_{g,n}|^2=\Bigg|{\kappa[\omega]^{(2n-1)^2}\over
\kappa[\phi^n]}{\phi^n_1\wedge\cdots\wedge\phi^n_{N_n}\over
(\omega_1\wedge\cdots\wedge\omega_g)^{c_n}}\Bigg|^2 \ ,
\label{modulosquaremu}\ee maps to the non-chiral analog \be
V_{n}(\tau)={K_1^{(2n-1)^2}\over
K_n}{1\over(\det\tau_2)^{2n(n-1)}} \ ,
\label{eccoqua}\ee
which is a $(0,0)$-form on $\M_g$.
Equation (\ref{effennea}) shows that the building
blocks of $V_{n}$ are just the vector-valued Teichm\"uller
modular forms introduced in \cite{Matone:2011ic} \be
[i_{N_n+1},\ldots,i_{M_n}|\tau]=\epsilon_{i_1,\ldots,i_{M_n}}{\kappa[\omega^{(n)}_{i_1},\ldots,\omega^{(n)}_{i_{N_n}}]\over\kappa[\omega]^{(2n-1)^2}}
\ , \label{uarmo}\ee $i_1,\ldots,i_{M_n}\in\{1,\ldots,M_{n}\}$, and
that define the string measures \cite{Matone:2012wy}.

\noindent Recall that $\I$ denotes the set of conformal weights
$k\in\QQ$ and  $\J$ is the set of $n_k\in\ZZ/2$. We consider the
volume forms on $\M_g$ \be V[\J]=\Z_{Pol}\prod_{k\in\I} V_{k}^{n_k}
\ . \label{volumiii}\ee Note that since
$\kappa[\omega^{(n)}_{i_1},\ldots,\omega^{(n)}_{i_{N_n}}]$ vanishes
on the hyperelliptic loci with $g\geq3$ \cite{Matone:2011ic}, by
(\ref{effennea}) also $K_n$ vanishes there. Therefore, $V[\J]$,
depending on the set $\J$, may be vanishing or singular in such
loci. Consider
$$
V[-1_2]=K_2(\tau_2^{-1})\Bigg|{d\tau_{i_1}\wedge\cdots \wedge d\tau_{i_{3g-3}}\over
  \kappa[\omega^{(2)}_{i_1},\ldots,\omega^{(2)}_{i_{3g-3}}]}\Bigg|^2 \ .
$$
This can be very explicitly expressed up to $g=4$. For $g=2$ and $g=3$ we have
$$
V[-1_2]={|d\tau_{1}\wedge d\tau_2\wedge d\tau_3|^2\over (\det \tau_2)^3} \ ,
$$
$$
V[-1_2]={|d\tau_{1}\wedge\cdots\wedge d\tau_6|^2\over (\det \tau_2)^4} \ .
$$
In the case of genus four
\be
V[-1_2]=\sum_{ {i_{9}>\ldots>
i_1=1\atop j_{9}>\ldots
> j_{1}=1}}^{10} S_{4p}(\tau)
{|\tau_2^{-1}\tau_2^{-1}|^{i_1\ldots i_{9}}_{j_1\ldots
j_{9}}\over
\prod_{k=1}^{{9}}\chi_{i_k}\chi_{j_k}}\bar  S_{4q}(\tau)
 \Bigg|{d\tau_{1}\wedge\cdots\wedge
\widehat{d\tau_{k}}\wedge\cdots\wedge d\tau_{10}\over
S_{4k}(\tau)}\Bigg|^2 \ , \label{nicenessqua}\ee where
$p=I_{10}\backslash\{{i_1},\ldots,{i_{9}}\}$ and
$q=I_{10}\backslash\{{j_1},\ldots,{j_{9}}\}$ (note that here we are
using the single indexing notation introduced in the Appendix) and
$$S_{4ij}(Z)={1+\delta_{ij}\over 2}{\partial F_4(Z)\over \partial Z_{ij}} \ ,$$
with
$$F_g=2^g
\sum_{\delta\hbox{
even}}\theta^{16}[\delta](0,Z)-\bigl(\sum_{\delta\hbox{
even}}\theta^{8}[\delta](0,Z)\bigr)^2 \ .
$$
$F_4$ is the Schottky-Igusa form, and has the property of vanishing only on the Jacobian, so that
it provides the effective solution of the Schottky problem.
It is immediate to see that $V[-1_2]$ is the volume form
on $\M_g$ induced by the Siegel metric on the Siegel upper half-space. Its expression for any genus, but without the use of theta constants,
was given in \cite{Matone:1900zz} (see also \cite{Matone:2006bb} and \cite{Matone:2005bx}).

{}From the above findings it follows that
 \be V_n=Y_n\int DX  e^{-S[X]} \ , \label{marvaillaise}\ee where $S[X]$ is the Polyakov action in
$2c_n$ dimensions
and
\be
Y_n=\int Db D\bar b Dc D\bar
c{\prod_{i}b(z_i)\bar b(z_i)\over \det B^{\circ n}(z_j,\bar
z_k)} \exp(-{1\over2\pi}\int_C\sqrt g b\nabla^z_{1-n}c+c.c.) \ .
\label{ypsilonennne}\ee
By (\ref{nonchiraldetm}) and (\ref{kenneaaa}), it follows that
\be
Y_n={|\kappa[\phi^n]|^2\over K_n}{\det'\,\Delta_{1-n}\over\det\N_n } \ .
\label{yuyo}\ee
Furthermore, by (\ref{volumiii}) we have the
following Weyl anomaly free partition functions
\be \int_{\M_g}V[\J]=\int DgDXD\Psi \exp(-S[X]-S[\Psi]) \ ,
\label{formidabiledue}\ee where now $S[X]$ is the Polyakov action in
$D=26+2\sum_{k\in \I}n_kc_k$
dimensions.  $D\Psi$ is the product on $k\in \I$ of $|n_k|$
copies of the measure of weight $k$ $b$-$c$ systems, including the
zero modes insertion, if $n_k>0$, or $\beta$-$\gamma$ systems if
$n_k<0$. $S[\Psi]$ denotes the sum of the corresponding non-chiral
actions.

\subsection{Curvature forms}

Both $X_n$ in (\ref{nneerr}) and $Y_n$ in (\ref{yuyo}) provide an enumeration of the laplacian of determinants whose normalization eliminates the dependence
on the choice of the basis of $H^0(K_C^n)$. We saw that whereas $X_n$ naturally arises by looking for a Weyl and modular invariant determinant regularization, in the case of $Y_n$ the hyperelliptic loci may be
in their
divisor. Such a property, and the structure of both $X_n$ and $Y_n$, suggests that they represent key quantities to investigate the geometry of $\overline\M_g$.

Let us go back to the space $\overline{\cal M}_{g}$.
It turns out that the components ${D}_k$ of Deligne-Mumford boundary, introduced in (\ref{bobmarleys}), provide, together with
the divisor associated to Weil-Petersson class $[\omega_{WP}]/2\pi^2$, a basis for
$H_{6h-8}({\overline{\cal M}}_g,{\QQ})$.
Consider the universal curve
${\cal C}\overline{\cal M}_{g,n}$
 over $\overline{\cal M}_{g,n}$,
built by placing over each point of
$\overline{\cal M}_{g,n}$ the corresponding curve.
Note that $\overline{\cal M}_{g,1}$ can be identified
with ${\cal C}\overline{\cal M}_{g}$. More generally
$\overline{\cal M}_{g,n}$ can be identified with
${\cal C}_n\left(\overline {\cal M}_g\right)\backslash\{sing\}$
where ${\cal C}_n\left(\overline {\cal M}_g\right)$ denotes
 the $n$-fold fiber product of the
$n$ copies ${\cal C}_{(1)}\overline{\cal M}_g,
\ldots,{\cal C}_{(n)}\overline{\cal M}_g$
of the universal curve over $\overline{\cal M}_g$ and $\{sing\}$ is the
locus of ${\cal C}_n\left(\overline {\cal M}_g\right)$ where the punctures
come together.

Denote by $K_{{\cal C}/{\cal M}}$ the cotangent
 bundle to the fibers of ${\cal C}\overline {\cal M}_{g,n}
\to\overline{\cal M}_{g,n}$,
built by taking all the spaces of
$(1,0)$-forms on the various $\Sigma$ and pasting
them together into a bundle over ${\cal C}
\overline{\cal M}_{g,n}$. Let $C$ be a curve in $\overline{\cal M}_{g,n}$.
Consider the cotangent space $T^*C_{|_{z_i}}$. It varies
holomorphically with $z_i$ giving a holomorphic line bundle
${\cal L}_{(i)}$ on $\overline{\cal M}_{g,n}$.
Considering the $z_i$ as sections of the universal
 curve ${\cal C}\overline{\cal M}_{g,n}$
 we have  ${\cal L}_{(i)}=z_i^*\left(K_{{\cal C}/
{\cal M}}\right)$.

Let us consider the Witten intersection numbers \cite{Witten:1989ig}
 \be
\langle\tau_{d_1}\cdots\tau_{d_n}\rangle=
\int_{\overline{\cal M}_{g,n}}
c_1\left({\cal L}_{(1)}\right)^{d_1}\wedge\cdots
\wedge c_1\left({\cal L}_{(n)}
\right)^{d_n} \ ,
\label{43}\ee
which
are non-vanishing only if
$\sum d_i=3g-3+n$.
These are related to the
Mumford tautological classes \cite{Dmumford}
\be
\kappa_l=\pi_*\left(c_1\left({\cal L}\right)^{l+1}\right)
 =\int_{\pi^{-1}(p)}c_1\left({\cal L}\right)^{l+1} \ ,
\label{44}\ee
$p \in \overline {\cal M}_g$,
where $\cal L$ is the line bundle whose fiber is the
cotangent space to the one marked point of
$\overline{\cal M}_{g,1}$ and
$\pi : \overline {\cal M}_{g,1}\to\overline{\cal M}_g$
is the projection forgetting the puncture.
 The $\kappa$'s correlation functions
$\langle\kappa_{s_1}\cdots\kappa_{s_n}\rangle=
\langle\wedge_{i=1}^n\kappa_{s_i},\overline{\cal M}_g \rangle$,
 which are non-vanishing only if $\sum_i s_i=3g-3$,
are related to and $\tau$'s correlators.
 For example performing the integral over the fiber
of $\pi:\overline{\cal M}_{g,1}\to\overline{\cal M}_g$,
\be
\langle\tau_{3g-2} \rangle=
\int_{\overline{\cal M}_{g,1}}c_1({\cal L})^{3g-2}=
\int_{\overline{\cal M}_{g}}\kappa_{3g-3}=
\langle\kappa_{3g-3}\rangle \ .
\label{dhadlk}\ee
It is useful to express the $\kappa$'s correlators in the form \cite{Witten:1989ig}
 \be
\langle\kappa_{d_1-1}\cdots\kappa_{d_n-1}\rangle=
\int_{{\cal C}_n\left(\overline {\cal M}_g\right)}c_1
\left(\hat{\cal L}_{(1)}\right)^{d_1}
\wedge\cdots\wedge c_1\left(\hat{\cal L}_{(n)}
\right)^{d_n}   \ ,
\label{45}\ee
where $\hat{\cal L}_{(i)}=\pi_i^*\left(K_{{\cal C}_{(i)}/
{\cal M}}\right)$  and  $\pi_i:{\cal C}_n\left(\overline {\cal M}_g\right)
\to{\cal C}_{(i)}\overline{\cal M}_g$ is the natural projection.
Then notice that ${\cal C}_n\left(\overline {\cal M}_g\right)$
and $\overline {\cal M}_{g,n}$ differ for a divisor at infinity only.
This is the unique difference between
$\langle \kappa_{d_1-1}\cdots\kappa_{d_n-1}
\rangle$ and $\langle\tau_{d_1}\cdots\tau_{d_n}\rangle$. This leads to  relations for arbitrary correlators.

In \cite{wolpert3} Wolpert proved that the first tautological class
corresponds to the Weil-Petersson two-form
$$
\kappa_1=\omega_{WP}/\pi^2 \ .
$$
By  well-known results on $\omega_{WP}$, it follows that
\be
\kappa_1={6 i\over \pi }\overline\partial\partial \log {\det\,
\tau^{(2)}\over \det' \Delta_{\hat g,0}} \ ,
\label{ftclss}\ee
where $\Delta_{\hat g,0}$ denotes the laplacian with respect to the Poincar\'e metric and $\partial$,
 $\overline \partial$ are the holomorphic and antiholomorphic
components of the external derivative $d=\partial+\overline\partial$
on $\M_g$.
This implies that $\kappa_1$ can be seen as the curvature
form in ${\cal M}_g$
of the Hodge line bundle $(\lambda_1;\langle \,,\rangle_Q)$
endowed with the Quillen norm
\be
\langle \omega\,,\omega\rangle_Q=
{\det\,\tau^{(2)}\over \det' \Delta_{\hat g,0}} \ ,
\label{quillen}\ee
where $\omega\equiv\omega_1\wedge \cdots\wedge \omega_g$,
that is
\be
\kappa_1=12 c_1(\lambda_1) \ .
\label{didud}\ee
A natural question is whether this is a signal for the existence of more general relations between determinant of laplacians and the tautological classes. This would also imply
a relationship between 2D gravity, topological theories and the string determinants.

There is an obstruction to extend Eq.(\ref{didud}) to the tautological classes of higher degree. The reason is that in general the Quillen norm depends on the choice of the basis
of $H^0(K_C^n)$. The exception is just (\ref{quillen}) since in this case there is a canonical choice, that is $\omega_1,\ldots,\omega_g$, which is the one defining the
Hodge bundle. On the other hand, it has been shown in \cite{Matone:2011ic} that there are natural bases for $H^0(K_C^n)$ outside the hyperelliptic locus, the $\omega_i^{(n)}$'s. In the case $n=2$
this led to basic results on the Polyakov measure \cite{Matone:2012wy}, which in fact corresponds to the coefficients of the quadrics describing the world sheet in the projective space $\PP^{g-1}$.
This suggests defining
\be
\langle i_{N_n+1},\ldots,i_{M_n}\,, i_{N_n+1},\ldots,i_{M_n}\rangle= {\det\,\N_n(i_{N_n+1},\ldots,i_{M_n})\over \det' \Delta_{\hat g,1-n}} \ ,
\label{quillennew}\ee
with
$$
\det\N_n(i_{N_n+1},\ldots,i_{M_n})=\det \int_C\bar\omega_j^{(n)}{\hat\rho}^{1-n}\omega_k^{(n)} \ ,
$$
where $\hat\rho\equiv 2\hat g_{z\bar z}$ is the Poincar\'e metric tensor in local complex coordinates and the determinant is taken on the matrix's indices running
in the set $\{i_{1},\ldots,i_{N_n}\}$. Note that for $n=1$
(\ref{quillennew}) coincides with the Quillen norm. One may
immediately check that it holds \be |\mu_{g,n}|^2={\langle
\omega\,,\omega\rangle_Q^{c_n}\over\langle
i_{N_n+1},\ldots,i_{M_n}\,,
i_{N_n+1},\ldots,i_{M_n}\rangle}{|\omega_{i_{1}}^{(n)}\wedge\cdots\wedge
\omega_{i_{N_n}}^{(n)}|^2\over
|\omega_1\wedge\cdots\wedge\omega_g|^{2c_n}} \ .
\label{buonissima}\ee
On the other hand, the problem of the independence on the choice of the basis has been one of the main points of our initial investigation. This led us to introduce
$X_n$ and $Y_n$, which in fact do not depend on the choice of the basis $H^0(K^n_C)$. Therefore both $X_n$ and $Y_n$ provide an intrinsic way to define new curvature forms
\be
\sigma_n=c_1(X^{-1}_n[\hat g]) \ ,
\label{sigma}\ee
\be
\nu_n=c_1(Y^{-1}_n[\hat g]) \ ,
\label{nuu}\ee
where $X_n[\hat g]$ and $Y_n[\hat g]$ denote $X_n$ and $Y_n$ with $\Delta_{1-n}$ and $\N_n$ evaluated with respect to the Poincar\'e metric.
We conclude this section observing that related structures have been considered in \cite{Becchi:1995ik}.

\section{Further directions and conclusions}

We repeatedly saw that the key step in our construction concerns the
manifestly Weyl and modular invariant structure of the Bergman
reproducing kernel. It is constructed in terms of one of the
the bases of $H^0(K_C)$. We used it in two ways. In the first one $B^{1-n}(z,\bar
z)$ has been used to integrate the zero modes of the $b$ field of
weight $n$. Such a prescription is well defined on any Riemann
surface.  We also considered the determinant of the Hadamard
$n$-fold product of $B(z_i,\bar z_j)$. This is a modular invariant
quantity which is proportional to $|\det\phi_j(z_k)|^2$ but is
independent of the choice of the $\phi_1,\ldots,\phi_{N_n}$. As such
it provides the tool to absorb, in a modular invariant way, the
dependence on the points due to the insertion of the zero modes in
the path integral. Since $\det B^{\circ n}(z_i,\bar z_j)$ vanishes
on the hyperelliptic Riemann surfaces, it may happen that, depending
on the set $\J$, the resulting partition function vanishes or has
singularities there. Both $B^{1-n}(z,\bar z)$ and $\det B^{\circ
n}(z_i,\bar z_j)$ are related to the space of symmetric powers of
$H^0(K_C)$. The latter led to the concept of vector-valued
Teichm\"uller modular forms \cite{Matone:2011ic} which provide the
building blocks for the Mumford forms \cite{Matone:2011ic}. In
\cite{Matone:2012wy} it has been shown that such forms can be
expressed in terms of $K_n=M_n-N_n$ forms vanishing on the Jacobian,
thus extending to any genus the expression for the Polyakov measure
for $g=4$ conjectured by Belavin-Knizhink \cite{Belavin:1986cy} and
by Morozov \cite{MorozovDA}. This also suggested formulating the
bosonic string on the Siegel upper half-space, a matter related to
the problem of characterizing the Jacobian locus, i.e. the Schottky
problem. In \cite{Matone:2012wy} it was also shown that such
vector-valued Teichm\"uller modular forms appear in constructing the
superstring measure and in the Grushevsky ansatz
\cite{D'Hoker:2001zp} \cite{Matone:2005vm} \cite{Cacciatori:2007vk}
 \cite{Grushevsky:2008zm}
\cite{SalvatiManni:2008qa}-\cite{DuninBarkowski:2012ya}.

In  \cite{Matone:2005vm} it was suggested that the pure spinor
Berkovits formulation of superstring theory
\cite{BerkovitsFE}\cite{geometrica} may be related to the Schottky
problem. The reason is that the conditions of pure spinors are
reminiscent of the relations for the quadrics
$$
\sum_{i,j=1}^g C_{ij}^k\omega_i\omega_j=0 \ ,
$$
$k=1,\ldots,K_2$, describing $C$ in $\PP^{g-1}$. It has been shown
in \cite{Matone:2011ic} that the vector-valued Teichm\"uller modular
forms, i.e. the building blocks of the string measures, provide a
suitable combination of the coefficients of such quadrics. In
particular, it turns out that the
vector-valued Teichm\"uller modular forms are just the determinants
of such coefficients \cite{Matone:2012wy}.

We have seen that the Mumford forms relate
basic aspects in string theory, such as modular and Weyl invariance, to the geometry of $\M_g$. Until now, the unique
Mumford form of interest for string theories has been $\mu_{g,2}$, the one of the
bosonic string. On the other hand, we have seen that even the other
Mumford forms lead, by their non-chiral extension, to partition
functions which are volume forms on $\M_g$. In the Berkovits
construction, there are several fields leading to sections of
$\lambda_n$ and, due to the scalars, to powers of the Hodge bundle.
In general, the invariance under Weyl and modular transformations
provides strong constraints, in particular the one that follows from
metric integration requires the partition function to
be a volume form on $\M_g$. A further analysis of the Berkovits
approach may show a relation to the partition functions introduced
here. In this respect, it should be observed that the extension to
the case of fields with fractional weight, essentially reduces to the
problem of adding the dependence on the spin structures.

As we said, our construction is of interest also in superstring perturbation theories.
In this respect, let us recall that a considerable step in finding the superstring measure is the Grushevsky ansatz \cite{Grushevsky:2008zm}, which has been successful
in many respects. It satisfies quite stringent constraints up to
genus four. Recently, Dunin-Barkowski, Sleptsov and Stern proved
that the Grushevsky ansatz may satisfy such conditions up to genus five
\cite{DuninBarkowski:2012ya}. In \cite{Matone:2012wy} it has been
observed that it is natural to believe that the phenomenon
appearing at genus four, i.e. the Schottky-Igusa form
$F_4$ defines both the bosonic and superstring measures, generalizes
to higher genus. The reason is that since $F_4$ vanishes only on the
Jacobian and therefore characterizes it, one should expect that the
superstring measure, like the bosonic one \cite{Matone:2012wy},
continues to be characterized by the forms vanishing on the
Jacobian. Since these increase with the genus, they are
$K_2=M_2-N_2=(g-2)(g-3)/2$, one should expect that the extension of
Grushevsky's ansatz should involve all of them, not just one. In
particular, in genus five, one should expect three forms. Such an
observation is somehow related to the very interesting result by
Codogni and Shepherd-Barron, namely that it does not exist a stable
Schottky form \cite{CS}, so that, at least, one cannot expect that
the extension of Grushevsky's ansatz may involve only one form, this
should happen already at the genus five.

There is one more reason for that. One of the main results in recent
work on superstring perturbation theory \cite{Witten:2012bg} is
that, at least for $g\geq 5$, the moduli space of super Riemann
surfaces does not map to the moduli space of Riemann surfaces with a
spin structure. This result, and the appearance of more forms just
from $g=5$, may suggest the existence of some way to overcome the
problems in treating the super period matrix and related geometrical
quantities. In turn, this  may be related with the fact that
Grushevsky's ansatz involves fractional powers of forms that seem unlikely that could be well defined on the Jacobian with the increasing
of the genus.

A related aspect has been considered in
\cite{Bonora:1989wk} where  a
general method to absorb the spin fields in $b$-$c$ systems of real
weight was introduced. This may suggest considering a suitable extension of the
non-chiral analog of the Mumford forms to real weight. From
the point of view of the $b$-$c$ and $\beta$-$\gamma$ systems this
is always possible just because the action contains terms such as
$b\bar\partial c$ and $\beta\bar\partial \gamma$, which are well-defined $(1,1)$-forms even for $n\in\CC$.

Let us conclude by observing that some of the geometry underlying
the present construction has an interesting application to
Seiberg-Witten theory \cite{Seiberg:1994rs}\cite{Argyres:1994xh},
which will be considered elsewhere.

\section*{Acknowledgements}

I thank  Giulio Bonelli, Giulio Codogni, Maurizio Cornalba, Paolo Di
Vecchia, Pietro Grassi, Samuel Grushevsky, Ian Morrison, Paolo
Pasti, Augusto Sagnotti, Riccardo Salvati Manni, Dima Sorokin, Mario
Tonin and Roberto Volpato for helpful comments and discussions. This
research is supported by the Padova University Project CPDA119349
and by the MIUR-PRIN Contract No. 2009-KHZKRX.

\newpage

\appendix

\section{$\Sym^n H^0(K_C)$}

In the following we introduce a single indexing to denote quantities
such as $\omega_i\omega_j$, $i,j=1,\ldots,g$, by $\omega^{(2)}_i$,
$i=1,\ldots,g(g+1)/2$. More generally, one may consider the basis
$\tilde\omega_1^{(n)},\ldots,\tilde\omega_{M_n}^{(n)}$,
$M_n={g+n-1\choose n}$, of $\Sym^n H^0(K_C)$ whose elements are
symmetrized tensor products of $n$-tuples of vectors of the basis
$\omega_1,\ldots,\omega_g$, taken with respect to an arbitrary but
fixed ordering.  The image $\omega_i^{(n)}$, $i=1,\ldots, M_n$, of
$\tilde\omega_i^{(n)}$ under $\psi:\Sym^n H^0(K_C)\to H^0(K_C^n)$ is
surjective for $g=2$ and for $C$ non-hyperelliptic of genus $g>2$.
For each $n\in\ZZ_{>0}$, set $I_n=\{1,\ldots,n\}$.
Let us fix the index ordering and introduce some notation as in
\cite{Matone:2006bb}.

Let $V$ be a $g$-dimensional vector space and denote by
$$\Sym^nV\ni\eta_1\cdot\eta_2\cdots\eta_n=\sum_{s\in\perm_n}\eta_{s_1}\otimes\eta_{s_2}\otimes\ldots\otimes
\eta_{s_n}\ ,$$
the symmetrized tensor product of an $n$-tuple $(\eta_1,\ldots,\eta_n)$ of elements of $V$.
It is useful to fix an isomorphism $\CC^M\rightarrow\Sym^2\CC^g$ and, more
generally, an isomorphism $\CC^{M_n}\rightarrow\Sym^n\CC^g$, $n\in\ZZ_{>0}$.

Let $A:\CC^M\rightarrow{\rm Sym}^2\CC^g$, $M\equiv M_2$, be the isomorphism
$A(\tilde e_i)=\sprod{e_{\1_i}}{e_{\2_i}}$, with $\{\tilde
e_i\}_{i\in I_M}$ the canonical basis of $\CC^M$ and
$$(\1_i,\2_i)=\left\{\vcenter{\vbox{\halign{\strut\hskip 6pt $ # $ \hfil & \hskip 2cm$ # $ \hfil\cr (i,i)\ , &1\le i\le g\ ,\cr
(1,i-g+1)\ ,&g+1\le i\le 2g-1\ ,\cr(2,i-2g+3)\ , &2g\le i\le 3g-3\
,\cr \hfill\vdots \hfill& \hfill\vdots\hfill\cr (g-1,g)\
,&i=g(g+1)/2\ ,\cr}}}\right.$$ so that $\1_i\2_i$ is the $i$th
element in the $M$-tuple $(11,22,\ldots,gg,12,\ldots,1g,23,\ldots)$.
In general, one can define an
isomorphism $A:\CC^{M_n}\to {\rm Sym}^n\CC^g$, with $A(\tilde
e_i)=(e_{\1_i},\ldots,e_{\n_i})$, by fixing the $n$-tuples
$(\1_i,\ldots,\n_i)$, $i\in I_{M_n}$, in such a way that $\1_i\le
\2_i\le \ldots\le \n_i$.

Let $\perm_n$ be the group of permutations
of $n$ elements. For each vector $u=\tp(u_1,\ldots,u_g)\in\CC^g$ and matrix $A\in
M_g(\CC)$, set
$$\underbrace{u\cdots u_i}_{n\hbox{ times}}=\prod_{\m\in\{\1,\ldots,\n\}}u_{\m_i}\ ,\qquad
(\underbrace{A\cdots A}_{n\hbox{
times}})_{ij}=\sum_{s\in\perm_n}\prod_{\m\in\{\1,\ldots,\n\}}A_{\m_is(\m)_j}\
,$$ $i,j\in I_{M_n}$. In particular, let us define
$$\chi_i\equiv\chi_i^{(n)}=\prod_{k=1}^g\biggl(\sum_{\m\in\{\1,\ldots,\n\}}\delta_{k\m_i}\biggr)!=(\delta\cdots\delta)_{ii}\ ,$$ $i\in I_{M_n}$
(we will not write the superscript $(n)$ when it is clear from the
context), where $\delta$ denotes the identity matrix, so that, for
example,
$$\chi^{(2)}_i=1+\delta_{\1_i\2_i}\
,\qquad\quad
\chi_i^{(3)}=(1+\delta_{\1_i\2_i}+\delta_{\2_i\3_i})(1+\delta_{\1_i\3_i})\
.$$

\end{document}